\documentclass{icrc2009}

\usepackage{graphicx}   
\usepackage{caption}    
\usepackage{epsfig}    
\usepackage{fixltx2e}
\usepackage{url}

\newcommand{\shorttitle}[1]%
{\markboth{Proceedings of the 31\MakeLowercase{$^{st}$} ICRC, {\L}\'{o}d\'{z} 2009}{#1} }
\newcommand{\etal}{\MakeLowercase{\textit{et al. }}} 

\begin{document}
\title{Sky monitoring with ARGO-YBJ}

\author{\IEEEauthorblockN{S. Vernetto\IEEEauthorrefmark{1},
                          Z. Guglielmotto\IEEEauthorrefmark{2},
                          J.L. Zhang\IEEEauthorrefmark{3},
                          on behalf of the ARGO-YBJ Collaboration
                           }
                            \\
\IEEEauthorblockA{\IEEEauthorrefmark{1}IFSI-INAF and INFN Torino, Italy}
\IEEEauthorblockA{\IEEEauthorrefmark{2}
Dipartimento di Fisica Generale dell'Universit\`a di Torino, Torino, Italy}
\IEEEauthorblockA{\IEEEauthorrefmark{3}Institute of High Energy Physics, Chinese Academy of Science, Beijing, P.R. China}
}

\shorttitle{Vernetto \etal Sky monitoring with ARGO-YBJ}
\maketitle

\begin{abstract}

A sky monitoring at gamma ray energy E $>$ 0.6 TeV has been performed 
by the full coverage Extensive Air Shower detector ARGO-YBJ, located 
in Tibet at 4300 m of altitude.  
We monitored 135 galactic and extragalactic gamma 
ray sources in the sky declination band from -10$^{\circ}$ to 
+70$^{\circ}$ for 424 days, detecting
the Crab Nebula and Mrk421 with a significance respectively of 7.0 and
8.0 standard deviations.
For a set of 11 AGNs known to emit in the TeV energy range,
the search has been performed in time scales
of 1, 10 and 30 days in order to study possible flaring activities.
Significant emissions has been observed from Mrk421 
in the time scales of 10 and 30 days, during June and March 2008, when the 
source had a strong activity also observed in the X-rays waveband.

The analysis of the background has revealed the existence of a
significant excess of the CR flux in
two localized regions of angular size 10$^{\circ}$--30$^{\circ}$, 
in agreement with previous indications.

\end{abstract}

 
\section{The detector}

ARGO-YBJ is a full coverage air shower detector located at the
Yangbajing Cosmic Ray Laboratory 
(Tibet, P.R. China, lat 30.1N, long 90.5E, 4300 m a.s.l.).
Due to its large field of view ($\sim$2 sr) and its high
duty cycle ($>$90$\%$) it can monitor continously a large fraction of the sky,
searching for gamma rays sources at energy E $>$0.6 TeV.
 
The apparatus is composed of a central carpet of 
Resistive Plate Chambers (RPCs) 
($\sim$74$\times$ 78 m$^2$, $\sim$93$\%$ of active area)
enclosed by a guard ring with partial coverage, which allows to extend the
instrumented area up to
$\sim$100$\times$110 m$^2$. 
The basic data acquisition element is a cluster (5.7$\times$7.6
m$^2$), made of 12 RPCs (2.8$\times$1.25 m$^2$). Each
chamber is read by 80 strips of 6.75$\times$61.8 cm$^2$ (the
spatial pixel), logically organized in 10 independent pads of
55.6$\times$61.8 cm$^2$ which are individually acquired and
represent the time pixel of the detector. The full detector is
composed of 153 clusters for a total active surface of $\sim$6600
m$^2$ \cite{Aie06}. 

Operated in {\em shower mode} ARGO-YBJ records all events with a number 
of fired pads N$_{pad}\ge$ 20 in the central carpet,
detected in a time window of 420 ns. 
The spatial coordinates and the time of any
fired pad are then used to reconstruct the position of the shower
core and the arrival direction of the primary\cite{DiS07}.

 \begin{figure*}[ht]
  \centering
    \epsfig{file=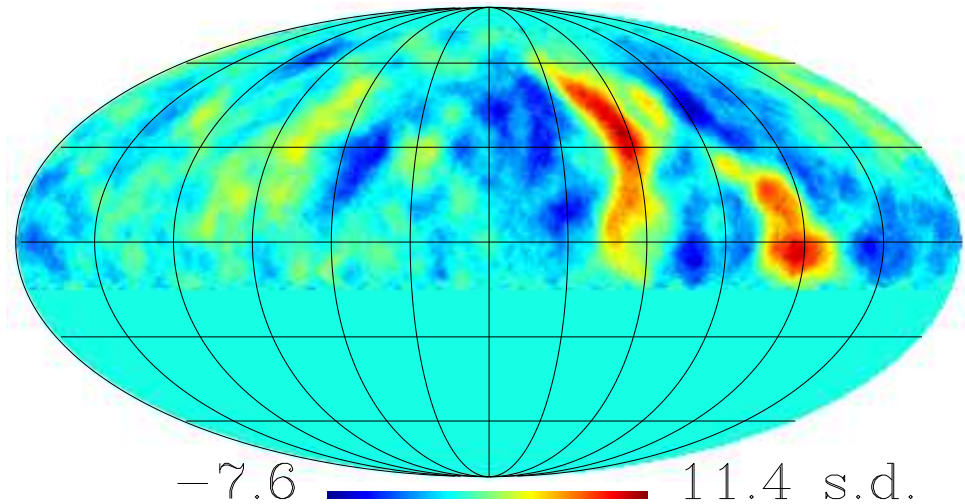,angle=90,width=5in}
  \caption{Sky map in equatorial coordinates 
    obtained in 424 days of measurements, for events
    with N$_{pad}\geq$40. The right ascension is equal to zero on the 
   extreme right of the map and increases going towards the left.
   The color scale indicates the statistical significance in 
   standard deviations.}
  \label{skymap}
 \end{figure*}

The installation of the detector has been completed in autumn
2007 and since November 2007 
ARGO-YBJ is in stable data taking with a trigger rate of $\sim$3.6 KHz.

The angular resolution and the pointing accuracy of the detector
have been evaluated by using the Moon shadow, i.e. the deficit of
cosmic rays in the Moon direction.\cite{luna}.
According to the Moon shadow data, the PSF of the detector is Gaussian for
N$_{pad} \ge$100, while for lower multiplicities it can be
described with an additional Gaussian, which contributes for about 20\%. 
The obtained values of the radius $\psi$ containing 
$\sim$71.5$\%$ of the signal, are 
2.59$^{\circ} \pm$ 0.16$^{\circ}$,  1.30$^{\circ} \pm$ 0.14 and  1.04$^{\circ} \pm$ 0.12$^{\circ}$ respectively 
for N$_{pad}\geq$40, 100 and 300,
in agreement with expectations from Monte Carlo simulations.
This measured angular resolution refers to cosmic rays-induced air
showers. The angular resolution for $\gamma$-induced events has
been evaluated by simulations and it is 10-30\% smaller
(depending on  N$_{pad}$),
due to the better defined time profile of the showers. 
 
\section{Sky monitoring}

The data used in this analysis have been recorded 
from 2007 day 311 to 2009 day 89
for a total live time of 424 days. 
All events with a zenith angle $\theta <$40$^{\circ}$  
and a number of hit pads on the central carpet
N$_{pad} \geq$40 have been considered.
The events with a large value of the $\chi^2$ resulting from the 
fitting procedure for the 
arrival direction determination, have been discarded from the
analysis.

With these events, five sky maps in celestial coordinates are built, 
for different number of hit pads: 
N$_{pad} \geq$40, 60, 100, 200, 300.
The maps are produced with the HEALPix package (Hierarchical Equal Area 
isoLatitude Pixelization)\cite{healpix}. Each map has $\sim$3.14 10$^6$ pixels
of equal area $\sim$0.013 squared degrees. 
The observed sky covers the declination band
from -10$^{\circ}$ to 70$^{\circ}$.

The background is evaluated with the {\em time swapping} 
method \cite{Ale92}. For each detected
event, 10 "fake" events are generated by replacing the original
arrival time with new ones, randomly selected from a buffer that
spans about 4 hours of data taking. 
For every N$_{pad}$ interval, the ``event'' map and the ``background map''
are then integrated over a
circular area of radius $\psi$, i.e. every bin is filled with the
content of all bins whose center has an angular distance less than
$\psi$ from its center.
The value of  the smoothing radius, related to the angular resolution, 
is the angle that maximizes the signal to noise ratio 
from a Crab-like source, according to simulations, 
i.e. $\psi$ = 1.3$^{\circ}$, 0.7$^{\circ}$ and  0.5$^{\circ}$ 
respectively for N$_{pad}\geq$40, 100 and 300.
Finally the background map is subtracted to the
relative event map, obtaining the "signal map",
where for every bin the statistical significance of the excess is
calculated.

 \begin{figure*}[ht]
  \centering
    \epsfig{file=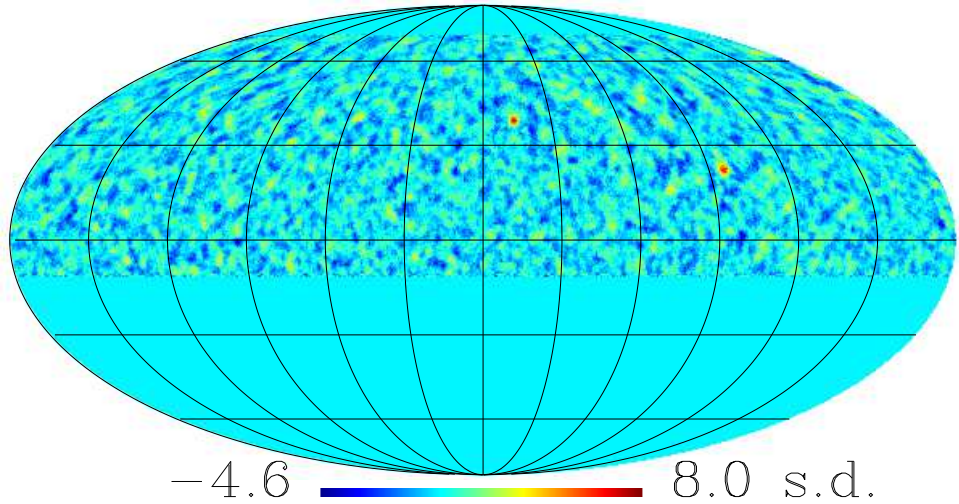,angle=90,width=5in}
  \caption{Sky map after correction, for events
    with N$_{pad}\geq$40.  The Crab Nebula and Mrk421 are observed
    with a statistical significance respectively of 7.0 and 8.0
    standard deviations.}
  \label{skymap2}
 \end{figure*}

Fig 1. shows the sky map for events with N$_{pad}\geq$40,
obtained with a smoothing radius $\psi$=5$^{\circ}$, adding the data
of the whole period. The map shows two large
hot spots in the region of the Galactic anticenter, 
already reported by the Milagro detector\cite{abdo}.
These regions have been interpreted as excesses of cosmic rays,
but the origin is not yet understood. They could be related
to the observed large scale anisotropy of cosmic rays 
due to the propagation of cosmic rays in the galactic magnetic field
\cite{anis1,anis2}.
Some authors discuss the possibility of the hot spots 
being due to a galactic nearby source as a
supernova explosion in the recent past\cite{ahar,salvati}.
The two excesses ($>$ 10 standard deviations, corresponding to
a flux increase of $\sim 0.1\%$) 
are observed by ARGO-YBJ around the positions 
$\alpha \sim$120$^{\circ}$, $\delta \sim$40$^{\circ}$ and
$\alpha \sim$60$^{\circ}$, $\delta \sim$-5$^{\circ}$.
in agreement with the Milagro detection, even if the maximum of the
second excess is slightly shifted
towards lower declinations, probably because 
ARGO-YBJ can observe
with more efficiency the southern regions of the sky,
being located at a smaller latitude than Milagro.
The two hot spots are also present in the map corresponding to events 
with a larger number of pad, even if with less significance 
due to the reduced statistics.
The median primary energy $E_m$ corresponding to proton showers
with N$_{pad}\geq$40 (300) is $\sim$ 2 (10) TeV. 

The deficit regions parallel to the excesses are due to
a known effect of the analysis, that uses also the excess events
to evaluate the background, artificially increasing the background.

 \begin{figure}[!t]
  \centering
  \includegraphics[width=2.5in]{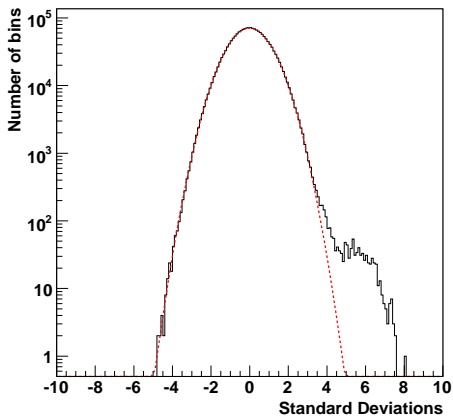}
  \caption{Distribution of the sky map excesses for
 events with N$_{pad}\geq$40. The red dashed line represents the Gaussian fit.}
 \end{figure}

The analysis of these hot spots will be the subject of a forthcoming paper.
In this work, we refer to them only since they affect the
point gamma ray source search. 
In fact the time swapping method applied over a time interval of 4 hours 
makes the analysis insensitive to 
features of scale larger than $\sim$60$^{\circ}$,
but all excesses of a smaller extension constitute a kind of ``noise''
for the point source search. If a source is located inside an excess region,
its significance will be overestimated, and vice versa.

In order to renormalized the background
eliminating all features of size  10-20$^{\circ}$,
we multiplied the content of each bin of the background map  
by a correction factor
f=E/B, where E and B are respectively the numbers of events of the
event map and background map inside a circle of radius 8$^{\circ}$ 
centered on that bin,
excluding the events from the circle central region  
up to a radius of 3.5$^{\circ}$ to avoid that a possible source
affects the evaluation of E.
Fig. 2 shows the sky map after the correction.
The Crab Nebula and Mrk421 are visible with a statistical
significance respectively of 7.0 and 8.0  standard deviations.
The obtained Crab energy spectrum is in agreement with the
observations by other experiments\cite{crab}.
Mrk421 had many active states
during 2008, in particular during March and June,
and has been deeply studied by ARGO-YBJ in a dedicate
paper\cite{m421}.

Excluding the bins around the Crab and Mrk421 positions, the distribution of 
the excesses well fits to a Gauss distribution
with mean value = -3.6$\pm$0.8 10$^{-3}$ and r.m.s.=1.01$\pm$ 5 10$^{-4}$ 
(Fig.3).

 \begin{figure}[!t]
  \centering
  \includegraphics[width=2.5in]{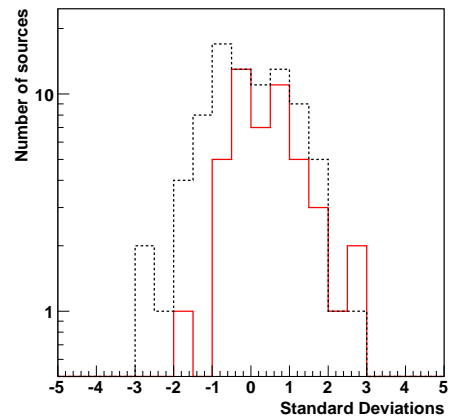}
  \caption{Distribution of the excesses observed at the position of
  133 gamma ray sources (black dashed line: extragalactic sources, red
  solid line: galactic sources)}
 \end{figure}

As a second step, we analised the maps contents at the positions of 
several known gamma rays sources.
We adopted two gamma ray source catalogues: 

a) the recently published 
list of bright sources observed by Fermi at E$>$100 MeV\cite{fermi} 

b) the list of sources seen at E$>$100 GeV\cite{tevcat} 

containing respectively 
115 and 44 sources in the declination band corresponding to the
ARGO-YBJ field of view, of which 24 are contained in both lists. 
Merging the two catalogues, we got
a list of 135 sources, 86 extragalactic and 49 galactic (or
unidentified but lying close to the galactic plane).
For every source the radius $r$ of the observational window is set to $\psi$, 
unless the source is known to be extended.
In this case $r$= $\sqrt(\psi^2$+$\alpha^2$) 
where $\alpha$ is equal to half the angular size.

  \begin{table}[th]
  \caption{AGNs observation: for each source is given 
   a) the redshift, 
   b) the culmination zenith angle $\theta_c$, 
   c) the hours of observation per day, 
   d) the flux above 0.6 TeV.}
  \vspace{+1pc}
  \centering
  \begin{tabular}{|c|c|c|c|c|}
  \hline
   Source  & $z$ & $\theta_c$ & h/day & Flux (E$>$0.6 Tev) \\
           &     &  (deg)     &      & ph cm$^{-2}$s$^{-1}$ \\
   \hline 
   M87          & 0.004 & 17.7 & 5.2 & $<$ 3.4 10$^{-11}$ \\
   Mrk421       & 0.031 & 8.1  & 6.4 & 5.5$\pm$0.7 10$^{-11}$   \\
   Mrk501       & 0.034 & 9.6  & 6.4 & $<$ 2.7 10$^{-11}$ \\
   BL Lacertae  & 0.069 & 12.2 & 6.4 & $<$ 1.7 10$^{-11}$ \\
   W Comae      & 0.102 & 1.9  & 6.1 & $<$ 2.4 10$^{-11}$ \\
   1H 1426+428  & 0.129 & 12.6 & 6.4 & $<$ 2.6 10$^{-11}$ \\
   1ES 0229+200 & 0.140 & 9.8  & 5.8 & $<$ 2.5 10$^{-11}$ \\
   1ES 1218+304 & 0.182 & 0.1  & 6.2 & $<$ 1.9 10$^{-11}$ \\
   1ES 1011+496 & 0.212 & 19.3 & 6.2 & $<$ 2.9 10$^{-11}$ \\
   PG 1553+113  & 0.360 & 18.9 & 5.1 & -------------- \\
   3C 66A       & 0.444 & 12.9 & 6.4 & -------------- \\
  \hline
  \end{tabular}
  \end{table}

The median energy $E_m$ corresponding to gamma rays showers 
with a given number of pads depends on the
source spectrum and on the culmination zenith angle.
For N$_{pad} \geq$40 the median energy ranges from $\sim$0.6 TeV (for
a source  culminating at the zenith and with a steep spectrum
like E$^{-3}$) up to $\sim$2 TeV (for a source culminating at
20$^{\circ}$ with a E$^{-2.5}$ spectrum).

Fig. 4 shows the statistical significance of the observed sources,
excluding the Crab Nebula and Mrk421, for N$_{pad} \geq$40.
No source shows a significance larger than 4 standard deviations,
however the mean value of the distribution of the excesses 
for galactic sources is clearly shifted towards positive values
(0.45$\pm$0.14) while the mean value 
for extragalactic sources is compatible with zero (-0.02$\pm$0.12).

We found 2 objects with a significance larger than 3 s.d for N$_{pad} \ge$60, 
namely:

a) MGROJ1908+06 with 3.2 s.d. (discovered by Milagro\cite{mila2}, confirmed
by Hess\cite{hess} and recently associated to the Fermi pulsar
0FGL J1907+5+0602)\cite{mila3};

b) the unidentified HESS source HESSJ1841-055, with 3.0 s.d.;

Considered the number of trials (133 sources $\times$5 N$_{pad}$ intervals) 
and the small significance
of the excesses it is difficult to say if they are due to gamma rays
or to a background fluctuation.
Moreover the corresponding flux is larger than what expected 
from these source, 
in particular for HESSJ1841-055\cite{hess2}.

\section{AGN follow up}

Since AGNs are known to be variable on different time scales, 
we have monitored the time behaviour of a subset of 11 AGNs known to
emit at E $>$ 100 GeV, and having a culmination zenith angle $<$20$^{\circ}$
in our field of view (see Tab.1).

The sources have been studied on time scales of 1, 10 days and 30 days.
For this analysis we have considered the data taken
in the period 2007 day 311 - 2009 day 89.

Concerning the daily search, we found only one excess with a
significance larger than 4 s.d. (4.3 s.d.), from the blazar 1ES0229+200 
on 2008 day 259, for  N$_{pad} \ge$ 40.
However, considering the number of trials
(11 sources $\times$510 days $\times$  5 N$_{pad}$ intervals)
this excess is consistent with a background fluctuation. 
 
Concerning the 10 days analysis, the only excess larger 
than 4 s.d. is due to Mrk421 in the time intervals 2008 days 161-170 
(4.6 s.d.) during a strong X-ray flare.
 
Looking for 30 days excess, the search has been done shifting
the 30 days interval in steps of 10 days. Also in this case we found several 
excesses from Mrk421 with significances between 4 and 5 s.d.,
in particular in the intervals: 2008 days 1-30, 71-100, 81-110,
91-120, 141-170, when several X-Ray flares have been observed\cite{rxte}.


For all sources except Mrk421 (for which we give the observed flux)
we calculate the upper limit to the flux at a confidence 
level of 3 standard deviations.
For each source we assumed a power law spectrum with a fixed
spectral slope: dN/dE = K E$^{-2.5}$ multiplied 
by an exponential factor  e$^{-\tau(E,z)}$
to take into account the absorption of gamma rays on 
the Extragalactic Background Light (EBL). 
We evaluated $\tau(E,z)$ interpolating
the curves given by Primack et al.\cite{Pri05} for fixed redshifts
in the range $z$=0.03-0.3. 
Lacking the absorption parameters outside this range, 
the upper limit for 3C66A ($z$=0.444) and PG1553+113 ($z$=0.36) 
have not been evaluated, while
for M87 ($z$=0.004) we assumed no absorption.

From the number of observed events (or the upper limit in case of no
detection) the corresponding gamma ray flux is determined  
by a complete simulation process that evaluates
the expected number of events from a source with a given spectrum
and a given daily path in the sky.
Tab.1 shows the observed integral flux above 0.6 TeV for Mrk421 
and the upper limits obtained for the 8 AGNs with no detection, averaged
over the whole observation period, for events with  N$_{pad}\geq$40. 

To know the minimum flux observable by ARGO-YBJ during a flare of  
a generic duration $n$ days,
one can multiply the upper limits given in the table by a factor 
$\sqrt(T/n)$, where $T$=424 is the total number
of days considered in this analysis.

\section{Conclusions}

For a period of $\sim$15 months ARGO-YBJ monitored
the gamma ray sky in the declination band from -10$^{\circ}$ to
+70$^{\circ}$.
The observations of the background  have revealed the  existence of
two regions  with size of order 10$^{\circ}$--30$^{\circ}$,
where the CR flux is enhanced  by  approximately 0.1 \%.
The origin of these  medium scale anisotropies,
previously observed  by the Milagro experiment, remains unexplained.

A search for point sources of high energy gamma  radiation
has resulted in the detection  with high significance of the Crab Nebula and Mrk421.

The data acquisition is currently going ahead with stable detection conditions. 
Studies to improve the detector sensitivity
are in progress, both in the direction of increasing the angular
resolution\cite{crab}
and of rejecting the cosmic rays background\cite{ghd}, implementing
gamma-hadron separation algorithms.

\end{document}